\documentclass[preprint,12pt]{elsarticle}

\usepackage{amssymb}
\usepackage{amsthm}
\usepackage{amsmath} 
\usepackage{amsfonts} 
\usepackage{stmaryrd}
\usepackage{url}
\usepackage{array}         
\usepackage{float}         
\usepackage[table]{xcolor} 
\usepackage{colortbl}
\usepackage{ulem}
\definecolor{darkgreen}{rgb}{0.0, 0.5, 0.0}

\journal{Automation in Construction}

\begin{document}

\begin{frontmatter}



\title{Assessment of ChatGPT for Engineering Statics Analysis}


\author[inst1]{Benjamin Hope}

\affiliation[inst1]{organization={Canterbury Fracture Group, Department of Civil and Natural Resources Engineering},
      addressline={University of Canterbury}, 
      city={Christchurch},
      country={New Zealand}}

\author[inst1]{Jayden Bracey}
\author[inst2]{Sahar Choukir}
\author[inst1,inst2]{Derek Warner}

\affiliation[inst2]{organization={Cornell Fracture Group, Civil and Environmental Engineering Department},
      addressline={Cornell University}, 
      city={Ithaca},
      postcode={14853}, 
      state={New York},
      country={United States}}

\begin{abstract}

Large language models (LLMs) such as OpenAI’s ChatGPT hold potential for automating engineering analysis, yet their reliability in solving multi-step statics problems remains uncertain. This study evaluates the performance of ChatGPT-4o and ChatGPT-o1-preview on foundational statics tasks, from simple $F=ma$ calculations to beam and truss analyses and compares their results to first-year engineering students on a typical statics exam. To enhance accuracy, we developed a Custom GPT, embedding refined prompts directly into its instructions. This optimized model achieved an 82\% score, surpassing the 75\% student average, demonstrating the impact of tailored guidance. Despite these improvements, LLMs continued to exhibit errors in nuanced or open-ended problems, such as misidentifying tension and compression in truss members. These findings highlight both the promise and current limitations of AI in structural analysis, emphasizing the need for improved reasoning, multimodal capabilities, and targeted training data for future AI-driven automation in civil and mechanical engineering.

\end{abstract}


\end{frontmatter}



\section{Introduction}
Generative AI models leverage machine learning (ML) to analyze and synthesize large datasets, enabling the generation of contextually relevant outputs across text, images, and audio domains. Among the most transformative advancements are large language models (LLMs) such as OpenAI’s ChatGPT, Anthropic’s Claude, and Google AI's Gemini. These tools are increasingly being utilized to tackle complex problems, 
simplify abstract concepts, automate routine calculations, and provide immediate, personalized assistance and feedback \cite{wei2022chain,joshi2023great,white2024preliminary,frenkel2024chatgpt,uddin2024chatgpt}. In the context of mechanics, LLMs can leverage not only classical theory but also numerical approaches such as the finite element method \cite{ni2024mechagents}.
On the other hand, the unregulated use of LLMs raises concerns 
about intellectual property, superficial engagement with core principles, and the potential erosion of critical thinking skills essential for professional engineering practice \cite{quibeldey2023disrupting,morsy2023challenges,zeeshan2024chatgpt,saad2024ai}.

Recent studies have examined the performance of LLMs on standardized tests such as the Medical College Admission Test (MCAT) and the Scholastic Aptitude Test (SAT), as well as their broader application in higher education \cite{bommineni2023performance,hickman2024performance,tan2024ai,kipp2024gpt,newton2024chatgpt}. Upon the public release of GPT-4, OpenAI evaluated its performance across several key program entry and professional licensure assessments, including the Graduate Record Examination (GRE), SAT, Law School Admission Test (LSAT), and the Uniform Bar Exam \cite{openai2023gpt}. In many instances, GPT-4 scored at or above the 90th percentile compared to human test-takers. However, it was generally found that while LLMs excelled in verbal reasoning, their performance in quantitative reasoning was notably weak \cite{hickman2024performance}. A study by Bommineni et al. \cite{bommineni2023performance} specifically assessed ChatGPT's performance on a 230-question MCAT, covering a wide range of topics in the social, physical, natural, and behavioral sciences, along with reasoning and critical analysis. The results revealed that ChatGPT performed at or above the median level of over 200,000 test-takers, with its responses and explanations closely aligning with the official answers. When tested on 30 unique questions not available online, GPT-4o maintained an impressive 96\% accuracy rate \cite{kipp2024gpt}. Furthermore, GPT-4o consistently outperformed medical students across six state exams, with a statistically significant mean score of 95.54\%, compared with the students’ 72.15\% \cite{kipp2024gpt}. 

This raises the question: how should we expect LLMs to perform on basic engineering questions and on engineering education assessments? As engineering problems often involve complex quantitative reasoning, the ability of models like GPT-4o to handle intricate calculations and multi-step solutions remains a critical factor for their potential application in this domain. Despite the widespread adoption of LLMs, their capabilities in solving engineering statics problems — a fundamental component of nearly all mechanics problems and of first-year civil and mechanical engineering curriculum — remain largely undocumented in the literature. This study seeks to addresses this gap by evaluating the performance of ChatGPT (ChatGPT-4o and o1-preview) across tasks of increasing complexity, from the basic application of Newton’s second law, to standard statics problems and culminating in a first-year engineering statics exam. We tested various prompting strategies, including zero-shot, few-shot, and chain-of-thought (CoT) techniques, and developed a Custom GPT incorporating optimized prompts. The study examined not only the models’ problem-solving capabilities but also their potential as automated tools in engineering education, providing insights into how they might assist students in mastering key concepts.

The remainder of this paper is organized as follows: Section 2 outlines the methodology, detailing the iterative approach to prompt engineering, problem selection, and evaluation metrics used to assess ChatGPT's performance. Section 3 examines ChatGPT’s ability to solve individual engineering statics problems in detail, including basic calculations based on Newton's second law of motion and analyses of beams and trusses, and evaluates the impact of text-based and image-based prompts on its performance. Section 4 presents ChatGPT’s performance on a first-year engineering statics exam, comparing it to student results and evaluating the performance of the Custom GPT designed for improved performance. Section 5 concludes with key insights on the role of AI in engineering mechanics education. 
\section{Methodology}
\subsection{ChatGPT Models}
ChatGPT, developed by OpenAI, is a one of the most recognized and one of the most widely used large language model (LLM). Since its release in late 2022, it has been central to the rise of generative AI, influencing the development of others LLMs with diverse capabilities. Our research focuses specifically  on evaluating the effectiveness of ChatGPT in engineering statics analysis.

Two versions of ChatGPT were assessed during this study: ChatGPT-4o and ChatGPT-o1-preview. ChatGPT-4o, noted for its multimodal capabilities, can process text and images, browse the internet, and support file handling. This versatility makes it a powerful tool for a wide range of applications, including those requiring multi-step problem-solving or interactive data analysis. Conversely, ChatGPT-o1-preview represents OpenAI's latest advancements in LLM technology, specifically optimized for advanced reasoning tasks. It is advertised as possessing PhD-level intelligence and is reported to exhibit reduced hallucination tendencies compared to its predecessor, ChatGPT-4o \cite{latif2024systematic}.

In its initial release, ChatGPT-o1-preview  was initially constrained by strict usage limits, allowing only 50 messages per week. This limited availability, combined with its higher cost, made ChatGPT-4o the primary model used for the majority of this study due to its greater accessibility, faster response times, and fewer restrictions during the research period. However, with the full release of ChatGPT-o1 on December 5, 2024, the usage limit was increased to 50 messages per day, and additional features, such as image uploads and analysis, were introduced.

\subsection{Temperature and Hallucination}

This study maintained ChatGPT’s default temperature setting of 0.7 to align with typical user experiences. The temperature setting, which controls the randomness of the model's output, is critical for balancing creativity and accuracy. However, one key observation was the variability in ChatGPT’s responses, even when identical prompts were used. This inconsistency stems from the non-zero temperature, which introduces randomness into the response generation process. While beneficial for open-ended tasks, this feature poses challenges in engineering applications where precision and repeatability are essential, particularly for problems with unique solutions.

Another significant challenge was hallucination, where the model generated incorrect or fabricated information. While hallucinations can occur even at zero temperature—though less frequently—they are more likely when the model lacks sufficient context or when the task requires reasoning beyond its training data.This issue directly impacts the reliability of ChatGPT in engineering statics analysis, as accurate and consistent outputs are essential. While careful prompt engineering can reduce hallucinations and improve repeatability, it cannot fully eliminate these issues. 

Overall, the study acknowledged these limitations and focused on optimizing prompts to mitigate some of the issues arising from temperature and hallucination, ensuring outputs were as reliable as possible for the engineering problems evaluated.

\subsection{Prompt Engineering}
Prompt engineering is the systematic construction of inputs to optimize AI-generated outputs(\cite{dang2022prompt}). It can be critical to the performance of LLMs. Studies have shown that well-structured prompts can significantly improve the model's performance across various tasks (\cite{schulhoff2024prompt}). This study evaluated different prompting strategies to assess ChatGPT’s ability to solve engineering statics problems and explored ways to enhance its performance, focusing on three key approaches: zero-shot, few-shot, and chain-of-thought (CoT) prompting.

Zero-shot, one-shot, and few-shot prompting involve providing the model with zero, one, or a few examples of how to solve a problem before asking it to tackle similar tasks. Research by Brown et al. (\cite{brown2020language}) demonstrated that LLMs like ChatGPT function as effective few-shot learners, with performance improving as the number of examples in the prompt increases. In this study, these techniques were used to evaluate how ChatGPT’s performance varied with different levels of guidance, revealing that including worked examples often improved the model's accuracy in solving statics problems.

Chain-of-thought (CoT) prompting further enhances performance by breaking down complex, multi-step problems into smaller, more manageable steps. This approach, shown by Wei et al. \cite{wei2022chain} to outperform standard prompting for mathematical reasoning tasks, was particularly effective in improving ChatGPT’s ability to tackle engineering problems requiring sequential reasoning.

Combining few-shot and chain-of-thought (CoT) prompting techniques has been shown to improve the reliability and accuracy of large language models like ChatGPT (\cite{schulhoff2024prompt}). In this study, these combined techniques were used to provide structured guidance and step-by-step reasoning for engineering statics problems. The integration of examples (few-shot) with explicit reasoning processes (CoT) aimed to improve performance by reducing errors arising from misinterpretation or inconsistent logic. Notably, ChatGPT’s performance did not consistently depend on the complexity of the problem but was influenced more by the clarity and structure of the prompts provided. These combined approaches were iteratively refined to maximize accuracy and consistency across a range of problems, regardless of their complexity.
\subsection{Evaluation of Text and Image-Based Prompts}

The study evaluated both text-based and image-based prompts. Text-based prompts involved detailed problem descriptions, while image-based prompts tested the models' ability to interpret diagrams and integrate visual information into their problem-solving processes. While ChatGPT-4o supported multimodal inputs, initial findings revealed challenges in interpreting complex graphics, prompting a primary focus on text-based scenarios.

\subsection{Engineering Statics Problem Set}

The study focused on foundational topics from first-year civil and mechanical engineering mechanics and statics courses, including:
\begin{itemize}
    \item \textbf{Equilibrium Problems:} Determining reaction forces for simply supported beams under various loads.
    \item \textbf{Truss Analysis:} Solving for forces in members of simple trusses subjected to point loads.
    \item \textbf{Beam Mechanics:} Analyzing bending moments, shear forces, and deflections.
    \item \textbf{First-Year Engineering Statics Test:} A combinations of problems mentioned above
\end{itemize}

These problems were selected for their significance in both engineering practice and education, aligning with key learning outcomes in international civil and mechanical engineering curricula.

\section{Engineering Mechanics Exemplary Problems}
In the following we will use the below format to present the prompts and ChatGPT outputs:
\begin{table}[H]
\centering
\begin{tabular}{|p{14cm}|}
\hline
\rowcolor{gray!20} 
\textbf{Prompt:} This box is a prompt to the ChatGPT. \\ \hline
\rowcolor{green!5}
\textbf{ChatGPT-4o Output:} Response from ChatGPT-4o. \textcolor{darkgreen}{Correct value highlighted green}, \textcolor{red}{incorrect value highlighted red}. \\ \hline
\rowcolor{blue!20}
\textbf{ChatGPT-o1-preview Output:} Response from ChatGPT-o1-preview. \textcolor{darkgreen}{Correct value highlighted green}, \textcolor{red}{incorrect value highlighted red}. \\ \hline
\end{tabular}
\label{tab:comparison}
\end{table}
\subsection{Basic Calculations}
To evaluate ChatGPT’s ability to perform the most basic mechanics calculation, a question requiring the application of Newton's second law was selected. At this stage, only the ChatGPT-4o model was tested. The evaluation began with the following prompt:

\begin{table}[H]
\centering
\begin{tabular}{|p{14cm}|}
\hline
\rowcolor{gray!20} 
\textbf{Prompt:} Provide answers to the following questions using the following equation. $F = m \cdot a$, where $F$ is force in newtons (N), $m$ is mass in kilograms (kg), and $a$ is acceleration in terms of gravity (g). \\ \hline
\end{tabular}
\label{tab:tab2}
\end{table}
Next specific questions were asked, requiring the calculation of force for a given mass ($m=4kg$) without specifying acceleration ($a$). This omission was intentional, designed to observe whether the model would recognize the missing information and either request it, assume a value, or proceed incorrectly.
\begin{table}[H]
\centering
\begin{tabular}{|p{14cm}|}
\hline
\rowcolor{gray!20} 
\textbf{Prompt:} Calculate $F$ for $m = 4$ kg. \\ \hline
\rowcolor{green!5}
\textbf{ChatGPT-4o Output:} In order to calculate the force $F$, we need the acceleration along with the mass $m$. Since you provided $m = 4$ kg, please provide the value of acceleration $a$. \\ \hline
\end{tabular}
\label{tab:tab3}
\end{table}

The missing information was correctly identified by ChatGPT, which then requested clarification. This was considered the appropriate and preferred response for the evaluation.

\begin{figure}[h]
    \centering
    \includegraphics[width=0.8\textwidth]{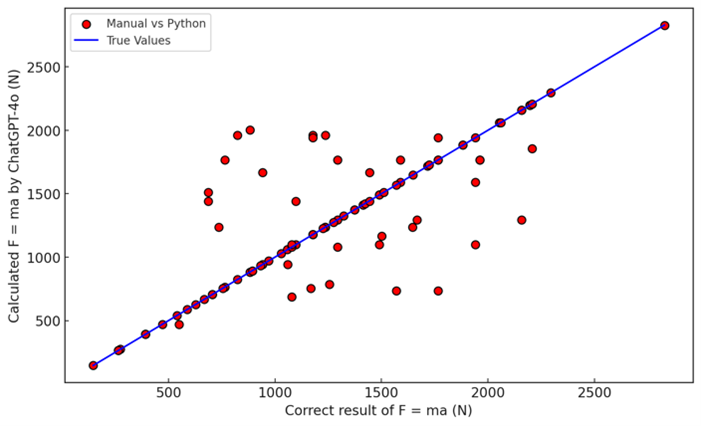}
    \caption{Accuracy of ChatGPT in calculating $F = m \cdot a$ across 100 iterations. The mass ($m$) and acceleration ($a$) values were randomly generated by ChatGPT for each calculation. The y-axis represents the force ($F$) values calculated by ChatGPT, and the x-axis represents the correct force values calculated using Python. The blue line represents perfect accuracy.}
    \label{fig:accuracy}
\end{figure}

Next, ChatGPT was instructed to calculate force over 100 iterations, with varying mass and acceleration values for each trial. The correct values were calculated using Python and compared to ChatGPT's results (Figure \ref{fig:accuracy}). The analysis revealed a correlation coefficient ($R^2$) of 0.79 between the ChatGPT outputs and the correct values, with a mean percentage difference of 13.73\%.

Upon investigating the discrepancies, it was discovered that ChatGPT performed the calculations using chain-of-thought (CoT) reasoning, breaking the problem down into smaller steps. For example, when $m = 4$ kg and $a = 10g$, ChatGPT produced the following steps:

\begin{table}[H]
\centering
\begin{tabular}{|p{14cm}|}
\hline
\rowcolor{green!5}
\textbf{ChatGPT-4o Output:} Multiply the mass by the acceleration: $10 \times 4 = 40$, Multiply $40$ by $9$: $40 \times 9 = 360$, Multiply $40$ by $0.81$: $40 \times 0.81 = 32.4$, Add the two results together: $360 + 32.4 = 392.4$. \\ \hline
\end{tabular}
\label{tab:tab4}
\end{table}

This process demonstrated that while ChatGPT correctly applied CoT reasoning, it introduced errors in the intermediate steps of multiplication. The unnecessary separation of 9.81 into 9 and 0.81 added complexity to the calculation. This breakdown introduced avoidable rounding and computational errors, compromising the accuracy of the final output across other trials.

The original prompt was repeated, explicitly instructing ChatGPT to use Python for all calculations. Under this approach, the model generated accurate results for every trial, effectively resolving the discrepancies. The instruction to use Python for all computations was maintained throughout the remainder of the research to ensure accurate calculations. This finding underscores the importance of clear and specific instructions when using LLMs for numerical tasks.

\subsection{Beams and Trusses} \label{problems}

This section examines ChatGPT’s ability to solve more complex engineering statics problems, including beam reactions and truss member forces. When tasked with simpler statics problems, such as single-force or single-support scenarios, ChatGPT generally followed correct procedural steps. For instance, in truss analysis, it accurately applied equilibrium equations to determine reactions before calculating member forces via the method of joints. However, its accuracy declined as the complexity of the problems increased—an increase in complexity often driven by the number of unknowns and equations, multiple or angled loads, extended geometry (e.g., multi-span beams or various support types), and the need for deeper theoretical reasoning. In engineering statics analysis, these factors collectively define a problem’s complexity: having to decompose inclined forces into orthogonal components, keeping track of multiple support reactions, determining determinacy and stability of trusses, or recognizing zero-force members each adds layers of challenge. Consequently, when problems involved multiple forces or inclined angles—or required a thorough understanding of member behavior under tension and compression—ChatGPT’s reliability diminished, underscoring the model’s difficulty in handling multi-step equilibrium analyses with higher conceptual demands.

Beam and truss examples are discussed in detail in Sections \ref{sec:beam_example} and \ref{sec:truss_example}, respectively, as they both contained common issues and nuances that ChatGPT struggled to address:
\begin{itemize}
    \item Difficulty in correctly resolving angled forces or members into their horizontal and vertical components.
    \item Errors in determining whether members were in tension or compression and in properly accounting for the direction of vectors when applying static equilibrium equations.
    \item Instances of hallucination, where the model introduced non-existent forces while using the method of joints or sections. This issue occurred more frequently when using multi-modal prompts compared to text-based prompts.
\end{itemize}

\subsubsection{Beam Example}\label{sec:beam_example}\mbox{} \\
\begin{figure}[h]
    \centering
    \includegraphics[width=0.8\textwidth]{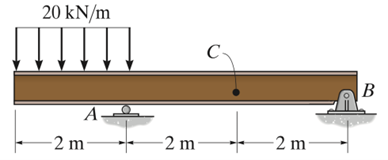}
    \caption{Beam problem presented to ChatGPT (\cite{hibbeler2018statics}).}
    \label{fig:beam_example}
\end{figure}
The following zero-shot prompt was first provided to the ChatGPT models to solve the beam problem shown in Figure \ref{fig:beam_example}:
\begin{table}[H]
\centering
\begin{tabular}{|p{14cm}|}
\hline
\rowcolor{gray!20} 
\textbf{Prompt:}  I have a 6 m long beam. Point A has a roller support located at $x = 2$ m. Point B has a pinned support located at $x = 6$ m. Point C is located at $x = 4$ m. A $20$ kN/m UDL is applied between $x = 0$ and $x = 2$ m. Determine the reactions at the supports, and the resultant normal force, shear force, and bending moment at Point C in the beam. Use Python for all calculations. \\ \hline
\rowcolor{green!5}
\textbf{ChatGPT-4o Output:} Reaction at A:\textcolor{red}{$10$ kN}, Reaction at B: \textcolor{red}{$30$ kN}, Shear Force at C: \textcolor{darkgreen}{$10$ kN}, Bending Moment at C: \textcolor{darkgreen}{$-20$ kNm}, Normal Force at C: \textcolor{darkgreen}{$0$ kN}. \\ \hline
\rowcolor{blue!20}
\textbf{ChatGPT-o1-preview Output:} Reaction at A: \textcolor{red}{$30$ kN}, Reaction at B: \textcolor{red}{$10$ kN}, Horizontal Reaction at B: \textcolor{darkgreen}{$0$ kN}, Shear Force at C: \textcolor{red}{$-10$ kN}, Bending Moment at C: \textcolor{red}{$60$ kNm}, Normal Force at C: \textcolor{darkgreen}{$0$ kN}. \\ \hline
\end{tabular}
\label{tab:tab6}
\end{table}
Both models failed to correctly calculate the reactions due to incorrect assumptions about the problem's constraints. Due to the loading on the beam to the left of support A, the beam should rotate and bend about support A, causing a negative reaction at support B (holding the beam down). The correct reactions are $50$ kN at A and $-10$ kN at B. Interestingly, 4o correctly solved the forces at Point C, which appeared to result from a coincidence of following the correct analysis process despite the incorrect reactions. In contrast, o1-preview did not follow standard procedures for calculating shear forces and bending moments, instead breaking the beam into segments and summing the results, which led to incorrect forces.
\begin{table}[H]
\centering
\begin{tabular}{|p{14cm}|}
\hline
\rowcolor{gray!20} 
\textbf{Prompt:}  Break this problem down step by step. 1) Solve for the reactions at the supports. Consider how the applied forces will create bending and rotation around supports and how that might influence the reactions at the supports. 2) Solve for the forces at point C using standard analysis procedures. 3) Review your calculations and answers, check that signs (+ and -) have been correctly applied, and if your answers make logical sense (i.e. equilibrium is satisfied). \\ \hline
\end{tabular}
\label{tab:tab7}
\end{table}
4o was further tested with a one, two, and three-shot CoT prompt, where each provided example became progressively more similar to the original question, with the third shot being nearly identical except for a different load magnitude. 
\begin{table}[H]
\centering
\begin{tabular}{|p{14cm}|}
\hline
\rowcolor{gray!20} 
\textbf{Prompt:} \textbf{\underline{Q1:}}  I have a 10 m long beam, Point A has a pinned support at $x = 0$, Point B has a roller support at $x = 5 \, \mathrm{m}$, Point C is located at $x = 2 \, \mathrm{m}$. A $5 \, \mathrm{kN/m}$ UDL is applied between $x = 5$ and $x = 10$. Determine the reactions at the supports, and the resultant normal force, shear force, and bending moment at point C in the beam. \textbf{\underline{A1:}} First, identifying that due to the loading of the beam to the right of support B, the beam will rotate and bend around support B, causing uplift and therefore a downwards reaction at support A.Sum of moments around point a (ACW = positive): \( 0 = 5B_y - 7.5(5 \times 5) \), solving for \( B_y \): \( B_y = \frac{7.5 \times 25}{5} = 37.5 \, \text{kNm} \). Sum of forces in the vertical direction (up = positive): \( 0 = B_y - A_y - 5 \times 5 \), so \( A_y = B_y - 25 = 37.5 - 25 = 12.5 \, \text{kN} \). From the Shear Force Diagram (SFD) and Bending Moment Diagram (BMD): \( V_c = -12.5 \, \text{kN}, M_c = -25 \, \text{kNm}, N_c = 0 \). \newline
\textbf{\underline{Q2:}} A 5 m long beam with Point \( A \) as a pinned support at \( x = 5 \, \text{m} \), Point \( B \) as a roller support at \( x = 3 \, \text{m} \), and Point \( C \) located at \( x = 4 \, \text{m} \). A UDL of \( 2 \, \text{kN/m} \) is applied between \( x = 0 \) and \( x = 3 \, \text{m} \). \textbf{\underline{A2:}} First, by identifying that due to the loading of the beam to the left of support \( B \), the beam will rotate and bend about support \( B \), causing uplift to the right of support \( B \), and therefore the reaction at support \( A \) is negative, holding the beam down. Sum of moments around point \( B \) (ACW = positive): \( 0 = 1.5(2 \times 3) - 2 \cdot A_y \), so \( A_y = \frac{9}{2} = 4.5 \, \text{kNm} \). Sum of forces in the vertical direction (up = positive): \( 0 = B_y - A_y - 2 \times 3 \), so \( B_y = A_y + 6 = 4.5 + 6 = 10.5 \, \text{kN} \). From the Shear Force Diagram (SFD) and Bending Moment Diagram (BMD): \( V_c = 4.5 \, \text{kN}, M_c = -4.5 \, \text{kNm}, N_c = 0 \). \newline  \textbf{\underline{Q3:}} I have a 6 m long beam, Point \( A \) has a pinned support at \( x = 6 \, \text{m} \), Point \( B \) has a roller support at \( x = 2 \, \text{m} \), Point \( C \) is located at \( x = 4 \, \text{m} \). A \( 10 \, \text{kN/m} \) UDL is applied between \( x = 0 \) and \( x = 2 \, \text{m} \). Determine the reactions at the supports, and the resultant normal force, shear force, and bending moment at Point \( C \) in the beam. \textbf{\underline{A3:}} First, by inspection, the loading of the beam to the left of support \( B \) will make the beam want to rotate anticlockwise and bend around support \( B \). This will cause uplift at Point \( A \), and therefore the reaction at support \( A \) is negative, holding the beam down. Sum of moments around Point \( B \) (ACW = positive): \( 0 = 1(2 \times 10) - 4 \cdot A_y \implies A_y = \frac{20}{4} = 5 \, \text{kNm} \, (\text{acting downwards}) \). Sum of forces in the vertical direction (up = positive): \( 0 = B_y - A_y - 2 \times 10 \implies B_y = A_y + 20 = 5 + 20 = 25 \, \text{kN} \). Drawing the Shear Force Diagram (SFD) and Bending Moment Diagram (BMD), I was able to find that \( V_c = 5 \, \text{kN}, M_c = -20 \, \text{kNm}, N_c = 0 \). \newline \textbf{\underline{Q4:}} *original prompt with CoT*
\\ \hline
\end{tabular}
\label{tab:tab8}
\end{table}

ChatGPT-4o was unable to solve the reactions correctly, and even after being provided with a nearly identical example to the original prompt, it continued to produce incorrect results for the reactions. This suggests that ChatGPT-4o was not effectively learning or adapting from the provided inputs.

\subsubsection{Truss Example}\label{sec:truss_example}\mbox{} \\
\begin{figure}[h]
    \centering
    \includegraphics[width=0.8\textwidth]{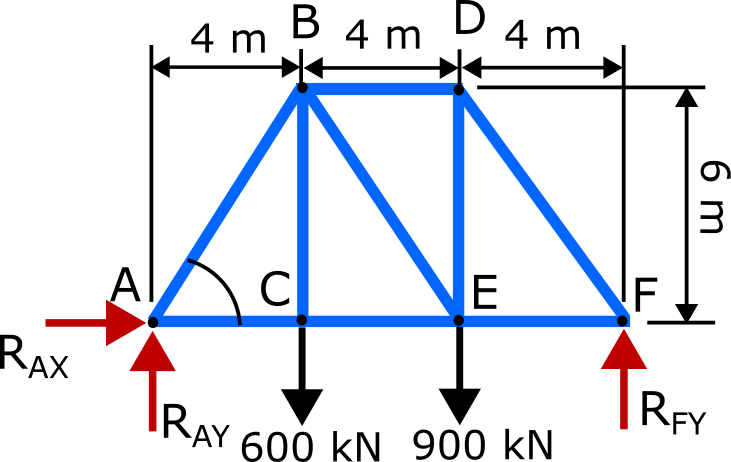}
    \caption{Truss problem presented to ChatGPT \cite{Fleischmann2019}.}
    \label{fig:truss_example}
\end{figure}
A zero-shot prompt describing the truss problem shown in  Figure \ref{fig:truss_example}  was first provided to the ChatGPT models:
\begin{table}[H]
\centering
\begin{tabular}{|p{14cm}|}
\hline
\rowcolor{gray!20} 
\textbf{Prompt:} I have a truss with the following nodes with (X, Y) coordinates: \textit{A: (0, 0)}, \textit{B: (4, 6)}, \textit{C: (4, 0)}, \textit{D: (8, 6)}, \textit{E: (8, 0)}, and \textit{F: (12, 0)}. Members connect nodes AB, AC, BC, BD, BE, CE, DE, DF, and EF. A $600$ kN force is applied in the vertical (Y) direction at node C, and a $900$ kN force is applied in the vertical (Y) direction at node E. The truss is supported by a pinned connection at node A, and a roller support at node F. Find the reactions at the supports and solve for the forces in each member, stating if it is in compression (C) or tension (T). Use Python for all calculations. \\ \hline
\rowcolor{green!5}
\textbf{ChatGPT-4o Output:} $R_{Ax}$ = \textcolor{darkgreen}{$0$ kN}, $R_{ay}$ =  \textcolor{darkgreen}{$700$ kN}, $R_{Fy}$ = \textcolor{darkgreen}{$800$ kN}, Member AB:  \textcolor{red}{$841.3$ kN} (\textcolor{red}{T}), Member AC: \textcolor{red}{$-466.7$} kN (\textcolor{red}{C}), Member BC: \textcolor{red}{$-600$ kN} (\textcolor{red}{C}), Member BD: \textcolor{darkgreen}{$-533.3$ kN} (\textcolor{darkgreen}{C}), Member BE: \textcolor{darkgreen}{$120.2$ kN} (\textcolor{darkgreen}{T}), Member CE: \textcolor{darkgreen}{$466.7$ kN} (\textcolor{darkgreen}{T}), Member DE: \textcolor{red}{$0$} (\textcolor{red}{No force}), Member DF: \textcolor{red}{$295.8$ kN} (\textcolor{red}{T}), Member EF: \textcolor{red}{$-44.4$ kN} (\textcolor{red}{C})\\ \hline
\rowcolor{blue!20}
\textbf{ChatGPT-o1-preview Output:} $R_{Ax}$ = \textcolor{darkgreen}{$0$ kN}, $R_{ay}$ = \textcolor{darkgreen}{$700$ kN}, \textcolor{darkgreen}{$R_{Fy}$ = $800$ kN}, Member AB: \textcolor{darkgreen}{$-841.63$ kN} (\textcolor{darkgreen}{C}), Member AC: \textcolor{red}{$467.89$ kN} (\textcolor{red}{T}), Member BC: \textcolor{darkgreen}{$600$ kN} (\textcolor{darkgreen}{T}), Member BE: \textcolor{red}{$-1081.06$ kN} (\textcolor{red}{C}), Member CE: \textcolor{red}{$-467.89$ kN} (\textcolor{red}{C}), Member EF: \textcolor{red}{$1067.89$ kN} (\textcolor{darkgreen}{T}), Member DF: \textcolor{red}{Cannot be accurately determined with this method}. BD and DE are likely \textcolor{red}{zero-force} members due to their configuration and the lack of loads at nodes B and D in the vertical direction. \\ \hline
\end{tabular}
\label{tab:tab9}
\end{table}
The correct answers for the truss problem are as follows: 
\begin{align*}
R_{AY} &= 700 \, \text{kN}, & F_{AB} &= -841.4 \, \text{kN} \, (\text{C}), & F_{CE} &= 466.8 \, \text{kN} \, (\text{T}), \\
F_{DF} &= -961.6 \, \text{kN} \, (\text{C}), & R_{AX} &= 0, & F_{AC} &= 466.8 \, \text{kN} \, (\text{T}), \\
F_{BE} &= 120.2 \, \text{kN} \, (\text{T}), & F_{DE} &= 800 \, \text{kN} \, (\text{T}), & R_{FY} &= 800 \, \text{kN}, \\
F_{BC} &= 600 \, \text{kN} \, (\text{T}), & F_{BD} &= -533.5 \, \text{kN} \, (\text{C}), & F_{EF} &= 533.5 \, \text{kN} \, (\text{T}).
\end{align*}
Both models successfully determined the reactions at the supports and calculated some member forces accurately; however, significant errors were observed in the majority of member force calculations. A recurring issue was the misidentification of whether members were in tension or compression, even when the numerical values of the forces were correct. This misclassification often propagated through the analysis, resulting in errors when summing forces at subsequent nodes. Despite these inaccuracies, both models demonstrated a fundamental understanding of the method of joints and adhered to its procedural framework.

Initially, both models attempted to write comprehensive Python code to solve the entire truss. This approach led to incorrect results, as the generated code failed to accurately account for the problem's nuances and constraints. To address this, the prompt was revised to clarify that Python should be used only for individual calculations rather than for automating the full truss analysis.

The amended prompt incorporated additional guidance, including detailed instructions, step-by-step examples, and chain-of-thought (CoT) reasoning techniques, to improve the models' performance and alignment with established engineering methodologies as shown below.

\begin{table}[H]
\centering
\begin{tabular}{|p{14cm}|}
\hline
\rowcolor{gray!20} 
\textbf{Prompt:} I have a truss with the following nodes…\newline …Write all formulas and equations out in full and use Python to solve completed equations, do not write a code to solve the full truss. Use the following convention: Forces pointing away from the node are tension and forces pointing towards the node are compression. When solving by sum of forces in x or y direction, consider the force sign relative to the node (example: if the force is in tension and coming from the left side of the node, it should be considered negative, but if a force is coming from the right side of the node, and is also in tension it should be considered positive. The same idea applies in the vertical). To start, assume all unknown forces are in tension (pointing away from the node). \newline When solving a truss, break the problem into small steps: first, solve for the reactions. Next, use the method of joints for each node by identifying the forces acting on a node and starting with nodes that have the least number of unknowns. For example, consider node \( A \): forces acting on node \( A \) include \( F_{AB} \), \( F_{AC} \), and \( R_{AY} \). Summing forces in the \( y \)-direction: \( R_{AY} + F_{AB}\sin(\theta) = 0 \implies F_{AB} = -\frac{R_{AY}}{\sin(\theta)} \). Since \( F_{AB} \) is negative, it points toward the node, indicating compression. Summing forces in the \( x \)-direction: \( F_{AC} - F_{AB}\cos(\theta) = 0 \implies F_{AC} = F_{AB}\cos(\theta) \). Since \( F_{AC} \) is positive, it points away from the node, confirming the initial tension assumption. \\ \hline
\end{tabular}
\label{tab:tab10}
\end{table}
With the inclusion of additional prompt details, o1-preview accurately calculated all force magnitudes but incorrectly identified that $F_{BC}$ was in compression. Similarly, the 4o-model exhibited improved performance yet  continued to produce errors in the force calculation, primarily due to misidentifying wether members were in tension and compression. Even aster providing both models with two- and three-shot examples for  node calculations, neither  as able to correctly classify $F_{BC}$  as being in tension. This issue persisted despite presenting  the complete formula and variable placeholders required for summing the vertical forces at node C, highlighting a limitation in the models’ ability to interpret and apply mechanical equilibrium principles accurately.

\subsection{Text-Based Prompt}
The challenges outlined in this study highlight that while ChatGPT can often be guided toward correct solutions, achieving correct outcomes typically requires multiple iterations, rendering the process less efficient than manual problem-solving. The implementation of few-shot and chain-of-thought (CoT) prompting techniques improved the models' accuracy. However, both models consistently exhibited difficulty in understanding the unique nuances of individual problems and demonstrated limited ability to retain and apply learned procedures across different tasks. This necessitated repeated intervention and guidance for each new problem.

Based on the trials conducted, it is evident that these models often make mistakes. This may stem from their reliance on pattern recognition, derived from training data, rather than employing deductive reasoning to systematically solve problems. It is possible that this indicates that LLMs may perform equally well on mechanics problems that are often perceived as more difficult by humans. For example, LLMs might do equally well on mechanics problems from the fourth year of a typically university engineering curriculum, provided sufficient training data. That said, problems perceived as more difficult do often require an increased number of steps and therefore could challenge LLMs in the same way that more complex beam and truss problems do. 

\subsection{Image-Based Prompts}

ChatGPT-4o’s ability to interpret image-based prompts was also evaluated, particularly with Truss diagrams to assess its performance in solving basic engineering statics problems.  Since mechanics problems often rely on visual aids like free-body diagrams to enhance comprehension, this evaluation examined whether providing prompts as images could improve ChatGPT’s accuracy and effectiveness, especially for visual learners.

\begin{figure}[h]
    \centering
    \includegraphics[width=0.8\textwidth]{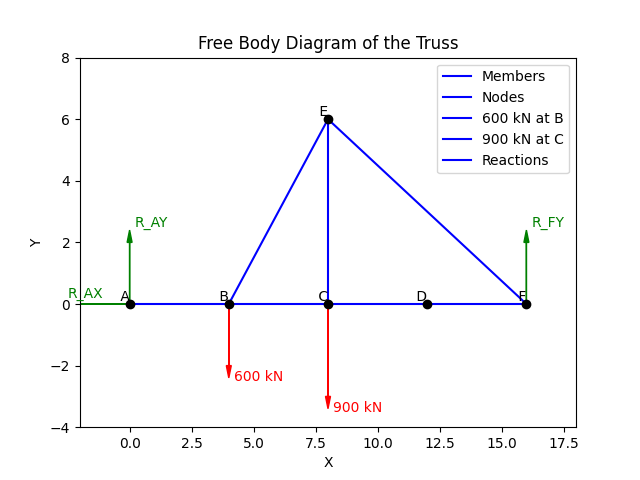}
    \caption{Incorrect diagram output produced by ChatGPT-4o from image-based prompt.}
    \label{fig:incorrect_diagram}
\end{figure}

Figure \ref{fig:truss_example} was presented with instructions to generate a free-body diagram, identify applied loads, and compute reactions at the supports. However, as shown in Figure \ref{fig:incorrect_diagram}, the model struggled with accurate image interpretation. Errors included incorrect node positioning, misrepresentation of dimensions, and the omission of key truss members (e.g., BD, BE, and DE), resulting in an incomplete and inaccurate depiction.

To examine these shortcomings, simplified hand-drawn sketches were subsequently utilized, omitting dimensions and text for clarity. While these modifications reduced some interpretative errors, ChatGPT-4o continued to struggle with more complex geometries and non-standard loading scenarios, such as triangular distributed loads or inclined members. Simpler problems, such as beams with uniform distributed loads or single point loads, were handled more effectively. These results suggest that while ChatGPT-4o can sometimes correctly interpret simple mechanics diagrams, it often struggles with more complex diagrams. 

\section{First-Year Engineering Statics Exam}\label{subsection2}

Although ChatGPT models (4o and o1-preview) make numerous errors in solving simple mechanics problems, this does not necessarily imply that their output is inferior to that of humans. To evaluate their problem-solving capabilities in comparison to human performance, we analyzed a set of elementary mechanics problems from a first-year engineering statics exam. While ChatGPT may not truly apply engineering statics concepts, it generates solutions by identifying patterns in its training data and extrapolating from similar examples. The exam was graded consistently for both students and ChatGPT models using a common rubric, with credit awarded for the correct identification and reproduction of key principles, regardless of whether the underlying reasoning was conceptual or pattern-based.

\begin{figure}[h!]
    \centering
    \includegraphics[width=\textwidth]{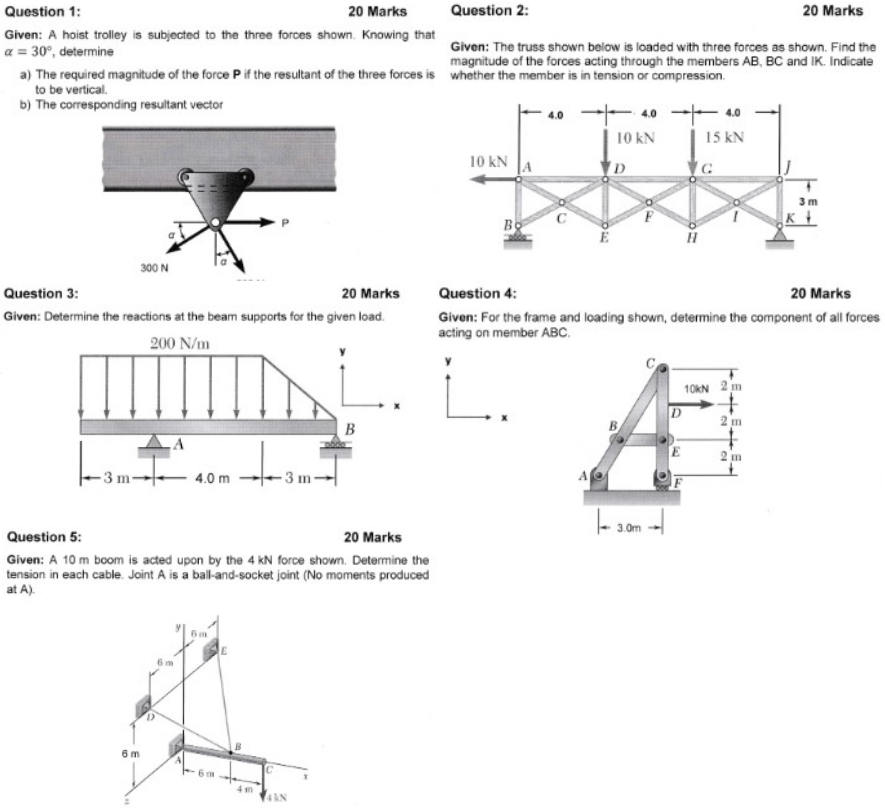}
    \caption{2024 First-Year Engineering Statics Exam referred to Exam 2. Problems adapted from \cite{beer2019vector,beer1999mechanics}.}
    \label{fig:exam}
\end{figure}
\subsection{Evaluation Setup}

Two exams were considered :

\begin{itemize}
    \item \textbf{Exam 1} served as a benchmark, enabling a comparison between student performance and ChatGPT’s 4o and o1-preview models. To establish a baseline for model evaluation, the Custom ChatGPT also attempted Exam 1. Student performance was based on the responses of 705 first-year engineering students.
    \item \textbf{Exam 2} contained similar problems and content to Exam 1 but included minor variations so that the correct answers were not identical. This exam was taken by 7 first-year engineering students and was used to assess the problem-solving skills of the 4o, o1-preview, and Custom ChatGPT models, as well as to examine the effects of different prompt styles.
\end{itemize}

Both exams were graded using a standardized rubric to ensure consistent evaluation of student and ChatGPT responses. This uniform assessment allowed meaningful comparisons between the models and human performance, despite the disparity in participant numbers for each exam.

To investigate the influence of prompt design, seven distinct prompt styles were tested, ranging from zero-shot to chain-of-thought reasoning, with some styles incorporating structured guidance for the models. These variations were intended to expose the models’ strengths (and potential weaknesses) across a range of problem-solving scenarios.

The seven distinct prompt styles are as follows:
\begin{enumerate}
    \item \textbf{Style 1:} Text-based questions (TBQs) without additional instructions (zero-shot).
    \item \textbf{Style 2:} TBQs with the instruction to "Talk me through every step and all of your thought processes," applying chain-of-thought (CoT) reasoning.
    \item \textbf{Style 3:} Image-based questions, tested exclusively on ChatGPT-4o.
    \item \textbf{Style 4:} TBQs with a specific note (Note 1) emphasizing attention to force directions, angles, and consistent sign conventions:
    \begin{itemize}
        \item Be very careful with directions of forces and angles, keep track of these during the entire calculations, and ensure your sign convention is applied correctly.
    \end{itemize}
    \item \textbf{Style 5:} TBQs combining CoT reasoning and Note 1 instructions.
    \item \textbf{Style 6:} Image-based questions incorporating CoT and Note 1, tested only on ChatGPT-4o.
    \item \textbf{Style 7:} Image-based questions with one-shot examples provided for each question, tested only on ChatGPT-4o.
\end{enumerate}

Exam 1, along with its marking rubric, was also provided to the Custom GPT model for use in solving Exam 2.

\begin{figure}[h!]
    \centering
    \includegraphics[width=\textwidth]{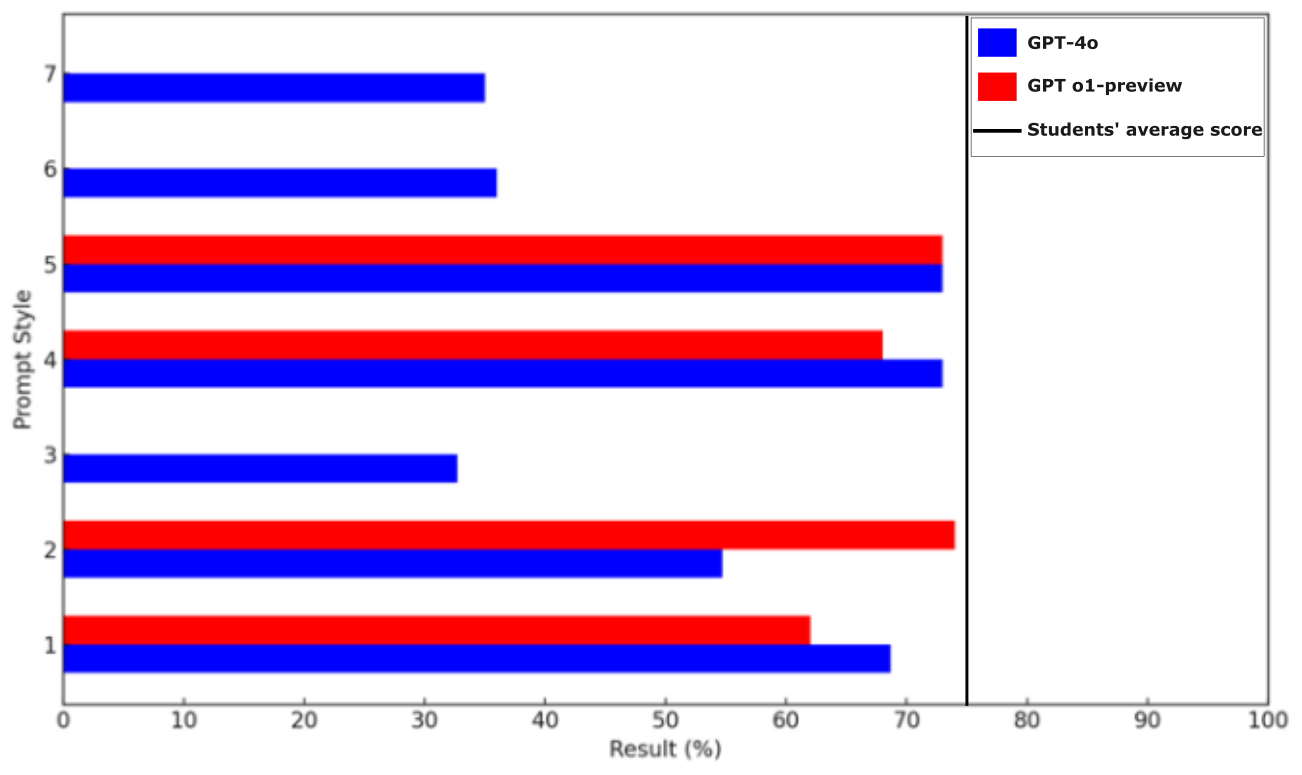}
    \caption{Results comparing different prompt styles used to evaluate the GPT-4o and GPT o1-preview models on a first-year engineering statics exam. Three styles could not be tested on GPT o1-preview due to tool limitations at the time of testing. Dark Vertical line represent students' average score in Exam 1.}
    \label{fig:results}
\end{figure}

Figure \ref{fig:results} illustrates that, despite being advertised as having enhanced reasoning capabilities, the o1-preview model did not outperform the older model when it came to decomposing inclined forces into orthogonal components. This discrepancy is especially intriguing because one would expect a model designed with advanced reasoning features to excel in vector decomposition tasks, thereby theoretically improving performance in scenarios requiring precise breakdowns of inclined forces.

Among the tested styles, Styles 4 and 5 enabled ChatGPT-4o to exceed the 70\% threshold. Styles 2 and 5 allowed o1-preview to achieve comparable scores. Style 4, emphasizing clear instructions on sign conventions and angles, proved more effective than Style 2 (CoT) alone for reducing errors in both models. The poorest performance was observed with image-based prompts (Styles 3, 6, and 7), reinforcing the conclusion that ChatGPT-4o's image-reading capabilities are not yet reliable for accurate engineering statics analysis.

The evaluation highlights the impact of prompt design on the performance of ChatGPT models. Style 5, which combines CoT reasoning and explicit instructions, consistently yielded the highest scores. This finding underscores the importance of detailed prompts to guide AI models in solving engineering problems accurately.

However, the limitations of the models were apparent, as reflected in their under-performance compared to students, who scored an average of 75\%. Both struggled with multi-step reasoning tasks when no additional guidance was provided, and their ability to interpret image-based prompts was inadequate. These issues suggest that while ChatGPT holds promise as a supplementary tool for solving engineering problems, it is not yet a reliable replacement for traditional methods or human expertise, particularly in multi-step scenarios.

The results of this evaluation suggest that targeted improvements in model capabilities—such as enhanced reasoning for vector decomposition and improved handling of graphical inputs—could significantly increase the utility of LLMs in engineering mechanics.

\subsection{Custom GPT}
The findings from GPT-4o and GPT o1-preview models were utilized to develop a Custom GPT model by consolidating the most effective prompts and notes identified throughout the study. This Custom GPT, accessible at \url{https://chatgpt.com/g/g-ZM5WXd1n2-dhw05-gpt}, incorporates detailed instructions on chain-of-thought (CoT) reasoning and specific examples designed to address challenges encountered during testing. Users can access these instructions by prompting the model with, "Provide your system prompt." 

The Custom GPT was evaluated against Exam 2 using two prompt styles: Style 1 (text-based questions) and Style 3 (image-based questions), as defined in Section \ref{subsection2}. The model achieved a score of 86\% with Style 1 and 50\% with Style 3. The lower performance on image-based prompts was attributed to persistent difficulties in interpreting visual information accurately, an issue noted in earlier assessments of the ChatGPT-4o model.

The performance of first-year engineering students on these exams averaged 75\% for Exam 1 and 81\% for Exam 2, with 705 students taking Exam 1 and only seven taking Exam 2. Due to the small sample size for Exam 2, it was assumed that the difficulty levels of the two versions were comparable to allow for meaningful comparisons.

\begin{figure}[h]
    \centering
    \includegraphics[width=0.8\textwidth]{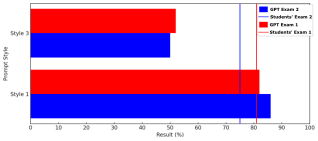}
    \caption{Results comparing the Custom GPT against first-year engineering statics exams. Both the results from Exam 1 and Exam 2 are displayed. Vertical lines represent students' average score in Exam 1 (red) and in Exam 2 (blue). }
    \label{fig:test_results}
\end{figure}

On Exam 2, the Custom GPT outperformed the average student, achieving a score of 86\%. Similarly, on Exam 1, the model scored 82\%, exceeding the class average of 75\% and placing it within the top 36\% of student results (Figure \ref{fig:test_results}). These results highlight the potential of the Custom GPT as a supplemental educational tool, particularly in text-based applications.

The Custom GPT demonstrated notable improvements over earlier models, particularly in correctly incorporating all relevant forces in equilibrium equations, accurately determining lever arms and applying correct sign conventions and recognizing that roller supports cannot provide horizontal reaction forces.

Despite these advances, several issues persisted.  In certain cases, the model incorrectly applied sign conventions, a common error among first-year students. The model occasionally assumed force directions without verification, leading to incorrect solutions. When prompted with Style 3 questions (image-based), the model struggled to process visual data effectively, often producing physically impossible or illogical outputs.

For instance, during Style 3 testing, the model incorrectly calculated a compressive force when solving for tension in a rope, failing to recognize the physical impossibility of the result. Furthermore, even when informed that the correct answer was included in its knowledge base, the model continued to produce erroneous results. This may be attributed to the prompt instructions being overly complex, potentially exceeding the model’s ability to process and respond efficiently.

Despite its improved performance on the statics Exam compared to ChatGPT-4o and ChatGPT-o1-preview, the Custom GPT often struggled with nuanced or open-ended problems. Persistent issues included misidentifying tension and compression in truss members and inconsistently handling inclined forces. While the model occasionally demonstrated elements of correct methodology, it rarely achieved fully accurate solutions. These limitations reflect that GPT models primarily rely on advanced pattern recognition derived from training data rather than systematically applying engineering principles.

\section{Conclusion}

This study assessed the capabilities of LLMs, particularly ChatGPT-4o, ChatGPT-o1-preview, and a Custom GPT model in solving engineering mechanics problems commonly encountered in first-year statics courses. By evaluating these models on progressively complex tasks, including simple mechanics problems including Newton's second law calculations ($F=ma$), truss and beam analyses, and a a first-year statics exam, we explored their strengths and limitations.

Key findings underscore the significant influence of prompt design on model performance. Chain-of-thought (CoT) reasoning and explicit instructions, such as those included in Style 5 prompts, consistently improved accuracy, with the Custom GPT model achieving the best results. However, challenges persist in handling nuanced or open-ended problems, such as misidentifying tension and compression in truss members or inconsistently resolving inclined forces. While the models demonstrated partial correctness and occasional adherence to engineering methodologies, they seldom produced fully accurate solutions. These results are consistent with LLMs reliance on pattern recognition derived from training data, rather than systematically applying engineering principles or deductive reasoning to solve problems.

The study also highlighted the limitations of ChatGPT-4o’s image-based capabilities, which failed to improve performance in engineering statics analysis tasks. While mechanics problems are often presented with diagrams and visual aids, ChatGPT-4o struggled to reliably interpret such prompts with the required accuracy. This limitation reinforces the need for enhanced multimodal capabilities to support visual problem-solving—a critical skill in engineering practice. Addressing these deficiencies could significantly broaden the applicability of AI tools in engineering analysis, design, and education. Targeted improvements in model capabilities—such as enhanced reasoning for vector decomposition and improved handling of graphical inputs—could significantly boost the utility of LLMs in engineering mechanics.

Interestingly, our study highlights a parallel in LLM performance with engineering students who often struggle with fundamental quantitative questions on exams. This raises a broader question: if students frequently make such errors, how do they perform robust engineering analysis and design in practice? The answer lies in the difference between academic assessments and real-world engineering practice. While exams emphasize theoretical understanding and the ability to perform under time constraints, real-world design is an iterative, collaborative process. Engineers rely on teamwork, peer reviews, simulations, prototypes, and built-in safety factors to ensure analyses and designs are reliable and effective. These processes allow errors to be identified and corrected, mitigating the impact of individual miscalculations. 

Similarly, while ChatGPT models currently struggle with fully mastering the theoretical rigor required for multi-step exam questions, their ability to provide rapid, partially correct solutions suggests potential as supplemental tools for iterative analysis and design processes. Like novice engineers learning on the job, these models could become more effective with training on domain-specific datasets, enhanced reasoning capabilities, multiagent modeling, and integration into collaborative workflows.

In summary, while ChatGPT is not yet a replacement for traditional learning or expert analysis, its ability to score comparably to students on engineering exams highlights its promise in shaping the future of engineering education and incorporation in engineering practice. Future iterations of these tools could bridge the gap between theoretical problem-solving and real-world application by incorporating modular instruction design, enhanced multimodal capabilities, and specialized training in engineering principles. By leveraging these advancements, LLMs could play a vital role in equipping the next generation of engineers with the tools they need to navigate the complexities of modern design and problem-solving.

\section*{Acknowledgments}
We would like to thank Hooman Pakdaman for his insightful contributions and active participation in this work. This research was supported by the National Science Foundation (1922081) and the New Zealand Ministry of Business, Innovation, and Employment (UOC2454). We also acknowledge the professional engineers who shared their expertise on AI applications in structural engineering, which helped refine our research objectives. Finally, we acknowledge the use of ChatGPT for its assistance in refining the language of this manuscript.

 \bibliographystyle{elsarticle-num} 
 \bibliography{cas-refs}





\end{document}